\documentclass[twocolumn]{aastex631}

\newcommand*{\Resize}[2]{\resizebox{#1}{!}{$#2$}}%

\newcommand{\alphaC}{\alpha_\mathrm{c}}
\newcommand{\alphaH}{\alpha_\mathrm{h}}

\newcommand{\Msun}{M_\odot}
\newcommand{\Mdot}{\dot{M}}
\newcommand{\Mdotc}{\dot{M}_\mathrm{c}}
\newcommand{\Mdotstar}{\dot{M}_\ast}
\newcommand{\Mdotin}{\dot{M}_\mathrm{in}}

\newcommand{\Mdotacc}{\dot{M}_\mathrm{acc}}

\newcommand{\masspro}{m_\mathrm{p}}
\newcommand{\muH}{\mu_\mathrm{h}}
\newcommand{\muC}{\mu_\mathrm{c}}

\newcommand{\rin}{r_\mathrm{in}}

\newcommand{\rco}{r_\mathrm{co}}

\newcommand{\rinmax}{r_\mathrm{in,max}}
\newcommand{\rcirc}{r_\mathrm{circ}}
\newcommand{\rA}{r_\mathrm{A}}
\newcommand{\reta}{r_\eta}
\newcommand{\rxi}{r_\xi}
\newcommand{\rh}{r_\mathrm{h}}

\newcommand{\rout}{r_\mathrm{out}}

\newcommand{\Rinmax}{R_\mathrm{in,max}}
\newcommand{\DeltaR}{\Delta r}
\newcommand{\rstar}{R_{*}}

\newcommand{\Ostar}{\Omega_\ast}
\newcommand{\OmegaK}{\Omega_\mathrm{K}}

\newcommand{\kb}{k_\mathrm{B}}

\newcommand{\Lacc}{L_\mathrm{acc}}
\newcommand{\Ldisc}{L_\mathrm{disc}}
\newcommand{\Ltot}{L_\mathrm{tot}}

\newcommand{\Lx}{L_\mathrm{X}}

\newcommand{\Firr}{F_\mathrm{irr}}

\newcommand{\tint}{\tau_\mathrm{int}}
\newcommand{\Tc}{T_\mathrm{c}}

\newcommand{\Teff}{T_\mathrm{eff}}
\newcommand{\Tirr}{T_\mathrm{irr}}
\newcommand{\Trec}{\tau_\mathrm{rec}}

\newcommand{\gpers}{g~s$^{-1}$}

\newcommand{\ergpers}{erg~s$^{-1}$}

\newcommand{\Sigmazero}{\Sigma_0}
\newcommand{\SigmaMin}{\Sigma_\mathrm{min}}
\newcommand{\SigmaMax}{\Sigma_\mathrm{max}}
\newcommand{\cs}{c_\mathrm{s}}

\newcommand{\Alfven}{Alfv$\acute{\mathrm{e}}$n~}

\newcommand{\x}{\;\times\;}

\begin{document}

\title{Typical X-ray Outburst Light Curves of Aql X-1\footnote{Accepted for publication in The Astrophysical Journal (ApJ)}}

\author[0000-0002-5887-6676]{Ömer Faruk \c{C}oban}
\affiliation{Faculty of Engineering and Natural Sciences, Sabancı University \\
Orhanlı, Tuzla 34956, İstanbul, Türkiye}

\author[0000-0002-1898-7902]{Ünal Ertan}
\affiliation{Faculty of Engineering and Natural Sciences, Sabancı University \\
Orhanlı, Tuzla 34956, İstanbul, Türkiye}



\begin{abstract}

We show that a typical X-ray outburst light curve of Aql X-1 can be reproduced by accretion onto the neutron star in the frame of the disk instability model without invoking partial accretion or propeller effect. The knee and the subsequent sharp decay in the X-ray light curve can be generated naturally by taking into account the weak dependence of the disk aspect ratio, $h/r$, on the disk mass-flow rate, $\Mdotin$, in the X-ray irradiation flux calculation. This $\Mdotin$ dependence of $h/r$ only slightly modifies the irradiation temperature profile along the hot disk in comparison to that obtained with constant $h/r$. Nevertheless, this small difference has a significant cumulative effect on the hot disk radius leading to a much faster decrease in the size of the hot disk, and thereby to a sharper decay in the X-ray outburst light curve. The same model also produces the long-term evolution of the source consistently with its observed outburst recurrence times and typical light curves of Aql X-1. Our results imply that the source accretes matter from the disk in the quiescent state as well. We also estimate that the dipole moment of the source $\mu \lesssim 4 \times 10^{26}$ G cm$^3~$($B \lesssim 4 \times 10^{8}$ G at the surface).

\end{abstract}



\section{Introduction}

Low-mass X-ray binaries (LMXBs) are systems containing either a neutron star (NS) or a black hole (BH) and a low-mass companion star with mass $M < 1\Msun$. 
In LMXBs, mass is transferred from the companion to the compact object through Roche-lobe overflow. Due to its angular momentum,  the matter cannot fall directly onto the star. Instead, the material forms a geometrically thin accretion disk around the NS \citep{FrankKing2002}. The matter in the disk moves in Keplerian orbits with speed $\upsilon_\mathrm{K} = (GM/r)^{1/2}$ where $G$ is the gravitational constant, $M$ is the mass of the NS, and $r$ is the radial distance from the center. Through viscous processes, angular momentum is carried outwards while the matter flows inwards toward the compact object. 

A small fraction of these systems are persistent X-ray sources, while the remaining larger fraction consists of transient systems, with both neutron stars (NSXTs) and black holes (BHXTs) that show X-ray outbursts with large ranges of recurrence and duration time-scales \citep[see e.g.][]{Psaltis2006}.  The inner disk in NSXTs is cut at a radius, $\rin$, depending on the mass-flow rate, $\Mdotin$, of the disk and the dipole moment $\mu = B \rstar^3$ of the NS, where $B$ and $\rstar$ are the dipole field strength at the equator and the radius of the star respectively. At the co-rotation radius $\rco = (GM/\Omega_*^2)^{1/3}$, the angular speed of the NS, $\Omega_*$, equals the Keplerian angular speed, $\OmegaK$, of the disk matter. If $\rin \leq \rco$, the matter can flow from the inner disk along the closed field lines onto the NS. In the case that $\rin > \rco$ the matter is estimated to be thrown out from the inner disk \citep[propeller effect;][]{Illarionov1975}. 
For a given $\mu$, the critical $\Mdotin$ level for the onset of the propeller mechanism is not well known. For spherical accretion onto a non-rotating magnetized star, magnetic pressure balances the ram pressure of the inflowing matter at \Alfven radius, $\rA \simeq (GM)^{-1/7} \mu^{4/7} \Mdotin^{-2/7}$ \citep{Davidson1973,Lamb1973}.  
In the case of disk accretion, $\rin$ is conventionally estimated by equating the viscous and magnetic stresses and found to be $\rxi = \xi \rA$ with $\sim 0.5 < \xi < 1$ \citep{GhoshLamb1979,kluzniak2007magnetically}.

It was proposed that there is a minimum critical X-ray luminosity, $\Lx$, for a persistent LMXB such that the entire disk is kept hot above the hydrogen ionization temperature by the X-ray irradiation flux with an $\Lx$ above this critical level \citep{vanParadijs1996}. When $\Lx$ decrease below the critical rate, the outer cold regions of the disk enter a low viscosity state with an inefficient mass transfer from the outer to the inner disk with a rate lower than the mass-flow rate, $\Mdotc$, from the companion to the outer disk. This leads to a gradual increase in the surface densities and temperatures of the outer disk. When the outer disk temperature exceeds the critical level at a radius, a heating front propagates to inner and outer radii, which could take most of the disk into the "hot state" with higher viscosities enhancing the mass transfer to the inner disk. The resultant abrupt increase in $\Mdotin$ at the inner disk is observed as an X-ray outburst. In the outburst phase, efficient mass transfer from the outer radii decreases the surface densities and temperatures of the outer disk. Eventually, starting from the outer disk, propagation of a cooling front inward could take the entire disk into the "cold state" with much lower $\Mdotin$ and $\Lx$ (quiescent state). To sum up, the outer disk behaves like a mass reservoir in the quiescent state, and this mass is released as a result of thermal-viscous disk instability producing an X-ray outburst (see e.g. \citealt{Lasota2001,FrankKing2002} for details).

The $\Lx$ peak in outbursts and the recurrence time-scale, $\Trec$, depend on the size of the disk together with the mass transfer rate from the companion, as well as the X-ray irradiation strength during the outburst. The X-ray irradiation slows down the inward propagation of the cooling front during the outburst state. Therefore, the outer disk of a system with relatively high $\Lx$ is evacuated more efficiently during outbursts, which needs a longer $\Trec$ to refill the disk. For instance, $\Trec$ is of the order of weeks for dwarf novae (DNs), while for NSXTs and BHXTs $\Trec$ varies from months to decades \citep{FrankKing2002}.

In the disk instability models (DIMs), the disk diffusion equation is usually solved using the $\alpha$-prescription of the kinematic viscosity $\nu = \alpha \cs h$~\citep{ShakuraSunyaev1973}, where $\cs$ is the sound speed and $h$ is the pressure scale-height of the disk. The basic characteristics of DN light curves can be reproduced with $\alphaH \simeq 0.1$ and $\alphaC \simeq 0.01-0.05$ for the hot and cold viscosity states, respectively. DIMs can also account for the outburst light curves and recurrence times of transient LMXBs with  $\alphaH$  and $\alphaC$ values similar to those used for DNs, provided that the effect of X-ray irradiation on the disk dynamics during the outburst states is included in the models \citep{king1998,KingRitter1998,Dubus1999,dubus2001}.

A characteristic "knee" followed by a steepening in the X-ray outburst light curves observed from a variety of magnetized stars in binaries including some NSXTs was proposed to be the signature of the onset of the propeller effect at a critical $\Lx$ corresponding to  $\rxi= \xi \rA = \rco$ with $\xi \simeq 0.5$ \citep{Zhang1998,Campana1998,Campana2018}. Among these NSXTs, Aql X-1 is a prototypical source with a spin period of $P = 1.8~\mathrm{ms}$ \citep{Casella2008}, accreting matter from a K-type companion. The source exhibits outbursts with a recurrence time $\Trec \sim 1$ yr. \citet{Gungor2014} classified these outbursts into three types based on their duration and peak flux. Among these, the 2010 outburst belongs to the most common type. During these outbursts, the $\Lx$ peak exceeds $10^{37}~$\ergpers~within a few days \citep{Campana2013}. After the peak, the X-ray light curve displays a smooth and slow exponential decay lasting for a few weeks, followed by a sharp turn downward, forming a knee-like morphology at an $\Lx$ level of a few $10^{36}~$\ergpers. Subsequently, $\Lx$ decreases more steeply to the quiescent level within several days \citep{Campana2014}. 

A steepening of the X-ray light curve during the decay phase is a behavior estimated in the DIM resulting from the inward propagation of the cooling front (see e.g. \citealt{KingRitter1998}). Recently, \citet{Lipunova2022} showed through detailed numerical analysis that the knee in the decay curve of Aql X-1 could be produced as a result of the shrinking size of the hot inner disk during the decay phase of the outburst. Initially, the X-ray light curve is similar to that of a purely viscous relaxation (without disk instability). The presence of a cold-hot border is communicated to the innermost disk after a viscous time across the hot inner disk. While $\rh$ is decreasing, the viscous time-scale of the hot disk also decreases. Below a critical level, this leads to a steeper $\Mdotin$ and $\Lx$ decay compared to those in purely viscous decay. This sequence of events governed by the dynamics and viscous time-scale of the inner hot disk causes a transition from a slower to a faster decay in $\Lx$, producing a knee-like morphology in the X-ray outburst light curve. \citet{Lipunova2022} modeled the X-ray outburst light curve of Aql X-1 with different assumptions (partial accretion, strong propeller, and total accretion). They obtain reasonable results with both the total accretion and partial accretion assumptions. For the total accretion case, they assumed that all the inflowing disk matter is accreted onto the star including the $\rin > \rco$ epoch. They calculate the total $\Lx$ including both accretion and disk luminosity, while $\Ldisc$ is dominating $\Lx$. During the decay phase, $\rin$ is increasing with decreasing $\Mdotin$. The effect of this outward propagation of the inner disk on the sharpening of the decay curve after the knee is not clear. \citet{Lipunova2022} pointed out that the $\Mdotin$ dependence of the $h/r$ ratio is significant in reproducing the sharper decay curve.

In this work, we also investigate the X-ray outburst light curves of Aql X-1 together with its recurrence characteristics in the frame of DIMs. In our model, the physical reason for the formation of the knee is similar to that found by \citet{Lipunova2022}. Taking the $\Mdotin$ dependence of the disk aspect ratio into account in the X-ray irradiation flux calculations, we demonstrate that the light curve morphology of Aql X-1 can be reproduced by the accretion luminosity without including partial accretion or a strong propeller in the model. As pointed out by \citet{Lipunova2022}, we also find that the weak $\Mdotin$ dependence of the disk aspect ratio significantly affects the decay morphology of the light curve. Furthermore, this model also reproduces the long-term characteristics of the recurrent outbursts of Aql X-1. The X-ray outburst light curves, the quiescent $\Lx$ level, and the recurrence time produced in the model are in good agreement with the observations.
In section \ref{sec:Model}, we describe our model briefly. We discuss the results of the model calculations in section \ref{sec:Results} and summarize our conclusions in section \ref{sec:Conclusions}.

\section{Model}
\label{sec:Model}

We use the numerical code employed earlier by \citet{Ertan2002} to simulate the fast-rise-exponential-decay (FRED) type outburst light curves of BHXTs. Here, we briefly describe the model calculations. 

We represent the mass distribution of the disk before the X-ray outburst with a Gaussian surface density distribution $\Sigma(r, t = 0) = \Sigmazero \exp \bigl[-\bigl\{(r-\rcirc/\DeltaR \bigl\}^2\bigr]$ centered at the circularisation radius, $\rcirc$. Through numerical calculations, we solve the diffusion equation 
\begin{equation}
\label{eqn:DiffEq}
\frac{\partial\Sigma}{\partial t} = \frac{3}{r}\frac{\partial}{\partial r} \left[ r^{1/2} \frac{\partial}{\partial r} (\nu \Sigma r^{1/2}) \right]
\end{equation}
\citep{FrankKing2002} to calculate $\Sigma (r,t)$ together with corresponding $\Mdotin$ evolution. We use the $\alpha$ prescription of the kinematic viscosity, $\nu = \alpha \cs h$ \citep{ShakuraSunyaev1973} where $h=\cs \OmegaK$ is the pressure scale height of the disk, $\cs=\sqrt{\kb \Tc / \mu \masspro}$ is the sound speed, $m_\mathrm{p}$ is the proton mass, $\mu$ is the mean molecular weight in units of $m_\mathrm{p}$, and  $\Tc$ is the mid-plane temperature of the disk. For a better resolution of the inner disk, we solve the diffusion equation substituting  $x = 2 r^{1/2}$ and $S = x \Sigma$ \citep{FrankKing2002}. We divide the disk into 800 radial grids equally spaced in $x$-space. Considering our assumption that all the inflowing disk matter is being accreted during the entire outburst phase, $\rin$ should be taken equal to or smaller than $\rco$ in the model. Nevertheless, the exact position of $\rin$ between $\rstar$ and $\rco=2.5\rstar$ does not affect $\Mdotin$ evolution estimated from the solution of equation~(\ref{eqn:DiffEq}). For the numerical calculations, we take $\rin=\rco$, and allow all the inflowing matter to flow through the inner boundary freely (We take $\Sigma=0$ at the inner disk edge). The outer disk is estimated to be cut by the strong tidal forces of the companion at a radius close to half of the Roche lobe radius \citep{FrankKing2002}, we set $\rout=1.8\times10^{11}~$cm, which is about $60\%$ of the Roche lobe radius. The exact value of $\rout$ does not affect the model light curve either. For the numerical calculations, we neglect the tidal torques and assume that the matter flowing freely outward across the outer disk edge is lost from the system (We set $\Sigma=0$ at $r = \rout$).

The viscous dissipation rate per unit disk area (from the mid-plane to the surface of the disk) can be written as
\begin{equation}
    \label{Eqn:Diss}
    D(r)=\frac{9}{8}\nu\Sigma\frac{GM}{r^3}=\frac{4\sigma}{3\tau}\Tc^4
\end{equation}
where $\sigma$ Stephan-Boltzmann constant, and $\tau=\kappa \Sigma$ is the vertically integrated optical depth of the disk. For an optically thick, geometrically thin disk, $\tau \gg 1$, and the dissipated energy 
is emitted locally in the form of blackbody radiation. 
We use the Rosseland mean opacities from the opacity tables for population I stars  (\citealt{Alexsander1994}, for $\log T \leq 3.7$ and \citealt{Iglesias1996}, for $\log T > 3.7$).

In addition to viscous dissipation, the disk is also heated by the X-rays produced by the accretion onto the NS. The X-ray irradiation flux for a point source is given by
\begin{equation}
    \label{Eqn:Firr}
    \Resize{7.7cm}{\Firr=\sigma \Tirr^4 = \frac{\eta \Mdotstar c^2 (1-\epsilon)}{4 \pi r^2}\frac{h}{r} \Bigl( \frac{\mathrm{d\:ln}h}{\mathrm{d\:ln}r}-1 \Bigr)= C \frac{\Mdotstar c^2}{4 \pi r^2}}
\end{equation}
\citep{ShakuraSunyaev1973,Dubus1999} where $\eta$ is the efficiency of the conversion of rest mass energy into radiation and $\epsilon$ is the X-ray albedo of the disk surface, and 
\begin{equation}
    \label{Eqn:C}
    C \equiv \eta (1-\epsilon)\frac{h}{r}\Biggl( \frac{\mathrm{d\:ln}h}{\mathrm{d\:ln}r}-1 \Biggr)
\end{equation}
is the irradiation parameter which is estimated to be in the range of $10^{-4} - 10^{-3}$ from the model fits to the optical and X-ray spectra of LMXBs \citep{deJong1996,Dubus1999,Ertan2002}. 
Note that if $\Firr$ is defined in terms of $\Lx = \eta \Mdotstar c^2$, then the irradiation parameter becomes greater than $C$ given by equation~(\ref{Eqn:C}) by a factor of $1 / \eta$. 

The effective temperature of the disk $\Teff \simeq \left[ (D + \Firr)/\sigma\right]^{1/4}$. At the inner disk regions, $D$ dominates $\Firr$, while outside a radius ($\sim10^{9}$ cm) the X-ray irradiation is the dominant heating mechanism. Since the $h/r$ ratio in the equation~(\ref{Eqn:C}) is not very sensitive to $\Mdotin$ and $r\:\bigl(h/r \propto \Mdotin^{3/20} r^{1/8}\bigr)$, it seems reasonable to take $C$ constant in the calculations \citep{FrankKing2002}. Nevertheless, in our simulations, we noticed that the variation of $h/r$ with $\Mdotin$ and the resultant change in $\Firr$ significantly modify the slope of the X-ray light curve in the sharp decay phase after the knee. Including the weak $\Mdotin$ and $r$ dependence of $h/r$, equation~(\ref{Eqn:C}) can be written as
\begin{equation}
    \label{C}
    C = C_0 \: \Mdot_{\mathrm{in},17}^{3/20} \: r_{10}^{1/8} 
\end{equation}
where $\Mdot_{\mathrm{in},17} = (\Mdotin / 10^{17} ~$\gpers), $~r_{10} = (r/10^{10}~\mathrm{cm})$, and $C_0$ will be a free parameter of our model. We note that \citet{Lipunova2022} considered this dependence of the irradiation parameter on $\Mdotin$ as well.

It was shown earlier through detailed analysis that the disk structure is not affected by the X-ray irradiation \citep{Dubus1999,Lasota2001}. This means that the irradiation flux does not change the $h/r$ ratio and its $\Mdotin$ dependence along the hot portion of the disk. Nevertheless, the irradiation significantly affects the stability criteria of the disk by reducing the minimum critical surface densities for the transition to the cold state \citep{king1998,KingRitter1998,Dubus1999,dubus2001,Lasota2001}.

At the peak of the X-ray outburst, a large portion of the disk is irradiated. With decreasing $\Lx$ in the decay phase, the radius of the hot disk, $r_h$, decreases as well. The irradiation flux, $\Firr$, determines the current position of the hot disk radius, $r_h$, and governs the speed of its inward propagation. In this phase, the X-ray light curve characteristics are produced mainly by the dynamics of the hot inner disk. Meanwhile, the outer cold regions of the disk that remain outside $r_h$ evolve with purely viscous decay, without any significant effect on the decay light curve. In other words, the $\Mdotin$ dependence of the disk aspect ratio enters the model calculations across the hot disk only which could be approximated by a steady disk profile corresponding to the current mass-flow rate of the disk.

Following the conventional approach, we calculate the kinematic viscosity with two different $\alpha$ parameters, $\alphaH$ and $\alphaC$, for the hot and cold viscosity states, respectively. For a given $r$ in the cold (hot) state, there is a maximum (minimum) critical surface density $\SigmaMax$ ($\SigmaMin$) for the transition to the hot (cold) state. We adopt $\SigmaMax$ and $\SigmaMin$ equations obtained by \citet{dubus2001} through detailed vertical disk analyses. Interpolating the critical $\Sigma$ values found for different radii of the disk, \citet{dubus2001} obtained
\begin{equation}
    \label{Eqn:SigmaMax}
    \Resize{7.7cm}{\SigmaMax = (10.8-10.3\: \xi)\: \alphaC^{-0.84} ~M_1^{-0.37+0.1 \xi} ~r_{10}^{1.11-0.27 \xi} ~ \mathrm{g\,cm^{-2}}}
\end{equation}
\begin{equation}
    \label{Eqn:SigmaMin}
    \Resize{6.6cm}{\SigmaMin = (8.3-7.1\: \xi)\: \alphaH^{-0.77} M_1^{-0.37} ~r_{10}^{1.12-0.23 \xi}~\mathrm{g\,cm^{-2}}}
\end{equation}
where $\xi = (\Tirr/10^4 ~\mathrm{K})^2$, and $M_1$ is the mass of the NS in solar masses. For our initial mass distribution, $\Lx \simeq 0$ until the surface densities increase at the inner disk, which is not realistic since $\Lx$ does not decrease below the quiescence level. Considering these situations, in addition to the critical surface densities, we also impose the $\alpha = \alphaH$ condition for radii at which $\Teff > 6000$ K, independent of $\Tirr$. 

The X-ray luminosity produced by accretion onto the star is related to the total disk luminosity through 
\begin{equation}
    \label{Eqn:Lacc}
    \Lacc \simeq \frac{GM\Mdotstar}{\rstar} \simeq \frac{2\rin}{\rstar} ~\Ldisc
\end{equation}
when $\rin$ is not close to $\rstar$. For the accretion onto the neutron star, $\rin$ should be either equal to or smaller than $\rco$. For Aql X-1, $\rco = 2.5 \times10^{6}~$cm. For the numerical calculations, we take $\rin=\rco$. Nevertheless, our results do not change for $\rin<\rco$. For instance, if $\rin=\rstar$, then the total X-ray luminosity $\Ltot = GM\Mdotstar/\rstar$ and half of which is produced by the disk, mostly from the innermost disk regions. Here, we neglect the effects due to the fast rotation of Aql X-1, which depends on $\rin$ (see e.g. \citealt{Lipunova2022}). The correction does not change our results by more than $25\%$ for both $\rin=\rstar$ and $\rin=\rco$ cases. For all our calculations, we take $M=1.4\Msun$ and $\rstar= 1\times10^{6}~$cm. During the transition to the propeller phase, $\Lx$ decreases rapidly by a factor greater than $\sim 2 \rco/\rstar \simeq 5$, since $\rin$ should be greater than $\rco$ in the propeller phase. Our results show that such a sharp modification of the model curve corresponding to accretion/propeller transition is not required to account for the observed outburst behavior of the source.

The maximum inner disk radius at which a steady propeller mechanism can be established could be written as
\begin{equation}
    \label{eqn:Rin} 	
	\Resize{7.7cm}{\Rinmax^{25/8}~\Big|1 - \Rinmax^{-3/2}\Big| ~ \simeq ~ 1.26  ~ \alpha_{-1}^{2/5} ~M_{1.4}^{-7/6} ~\Mdot_\mathrm{in,16}^{-7/20}~ \mu_{30}~P^{-13/12}}
\end{equation}
\citep{Ertan2017,Ertan2018}, where $\Rinmax=\rinmax/\rco$, $M_{1.4} = (M / 1.4 \Msun )$, $\Mdot_\mathrm{in,16} = ( \Mdotin / 10^{16}~$\gpers), $\mu_{30} = (\mu / 10^{30}$~G~cm$^3$), $\alpha_{-1} = (\alpha /0.1)$, and $P$ is the spin period of the NS. \citet{Ertan2017} derived this formula through analytical calculations adopting the basic results of the disk-magnetosphere interaction model proposed by \citet{lovelace1995}, which was developed later to simulate the propeller phase as well \citep{lovelace1999,ustyugova2006}. In this model, the inner disk settles down at a radius where the field lines can force the innermost disk matter into co-rotation. 
The inner disk-field interaction takes place in a narrow boundary with radial width $\Delta r < r$. The field lines interacting with matter inflate and open up within an interaction time-scale $\tint = \left|\OmegaK - \Ostar \right|^{-1}$. In the propeller phase, the matter can be thrown out along the open field lines. Subsequently, the lines reconnect on a time-scale similar to $\tint$. The field lines outside the boundary are open and decoupled from the disk. With these conditions, \citet{Ertan2017} found that the inflowing disk matter can be stopped at $\rA$, but cannot be thrown out from the system efficiently. The resultant pile-up pushes the inner disk inwards to the radius at which the field is strong enough to expel all the inflowing mass from the inner disk, continuously evacuating the inner boundary. Equation~(\ref{eqn:Rin}) gives the maximum radius at which this strong propeller condition is satisfied. In the propeller phase, $\rin = \reta$ does not scale with and in some cases significantly smaller than $\rA$ (see \citealt{Ertan2017,Ertan2018,Ertan2021} for details).  

The actual inner disk radius is estimated to be close to $\rinmax$, and can be written as $\reta=\eta \rinmax$ where $\eta$ is a parameter close to unity. The $\Lx$ level corresponding to strong propeller/accretion transition can be estimated from equation~(\ref{eqn:Rin}). The predictions of this model for the critical $\Lx$ and the torque variations are in agreement with the properties of transitional millisecond pulsars (tMSPs; see \citealt{Papitto2022} for a recent review) during their transitions between the LMXB and the radio millisecond pulsar (RMSP) states \citep{Ertan2017,Ertan2018}. The model was recently developed to include all the rotational phases of the NSs in LMXBs \citep{Ertan2018}, and applied to the torque-reversal properties of LMXBs as well \citep{gencali2022}. We use equation~\ref{eqn:Rin} to estimate the dipole field strength of Aql X-1 that is consistent with the results of our X-ray light curve analysis.  

We also perform simulations to analyze and compare the repeating outburst light curves produced in the model with the observed recurrent outbursts of Aql X-1. To represent the mass flow from the companion,  we add mass into the radial grid at $\rcirc$ at each time step. For Aql X-1, $\rco \simeq 2.5 \times 10^{6}~\mathrm{cm}$, $\rcirc = 5.0 \times 10^{10}~\mathrm{cm}$, and we take $\rout = 1.8 \times 10^{11}~\mathrm{cm}$ in our calculations. We have repeated the numerical calculations with different mass-flow rates, $\Mdotc$,  from the companion until we obtain typical outburst morphology and recurrence time of Aql X-1. In section \ref{sec:Results}, we discuss the results of the model calculations summarized above.

\section{Results and Discussion}
\label{sec:Results}

\subsection{The 2010 Outburst of Aql X-1}
\label{sec:Outburstfit}
\begin{figure}[ht!]
\centering
\includegraphics[width=\columnwidth]{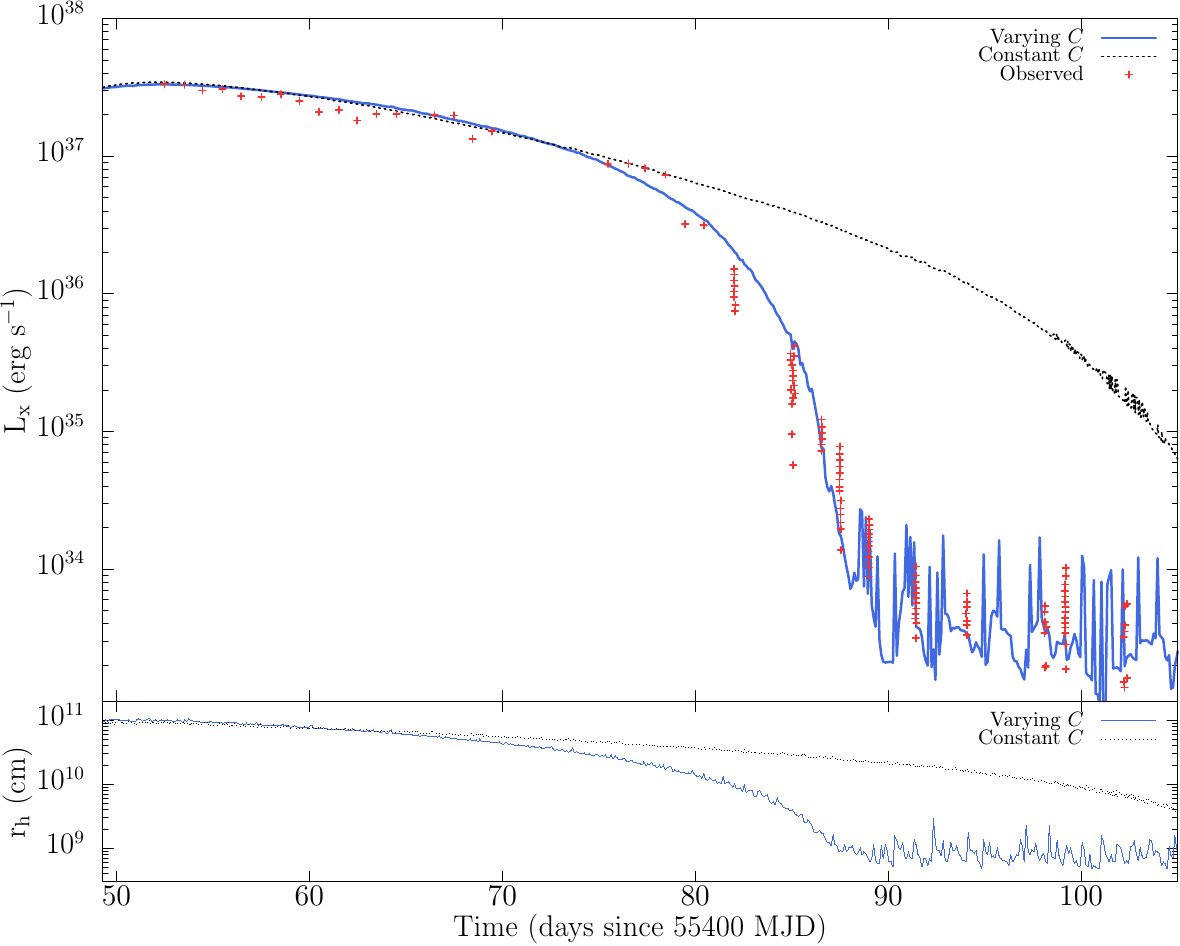}
\caption{X-ray outburst light curve of Aql X-1 taken from \citet{Campana2014}. The solid model curve is obtained with $\alphaC = 5\times10^{-3}, \alphaH = 0.1, \muC = 0.9, \muH =0.6$, and $C_0 = 1.3 \times 10^{-4}$. For the $\Lx$ calculation, we take $\Mdotstar = \Mdotin$ for the entire model curve. For comparison, the dotted curve is obtained with the same parameters, but with a constant irradiation parameter $C = 1.3 \times 10^{-4}$. The bottom panel shows the evolution of $\rh$ (blue curve). For comparison $\rh'$ (black curve) is also given.}
\label{fig:1}
\end{figure}

The model curve in Figure~\ref{fig:1} is obtained with $\alphaH=0.1$, $\muH=0.6$ in the hot state, and $\alphaC=5\times10^{-3}$, $\muC=0.9$ in the cold state. The model can reproduce the basic features of the observed X-ray light curve with $C_0 \simeq 1.3 \times 10^{-4}$. The fluctuations seen in the decay phase are not addressed in our model. Therefore, a seemingly good fit to data is sufficient for our purpose in this work. We note that the model curve represents the bolometric luminosity, while the observed luminosity is estimated from the $0.5-10~$keV flux. As the temperature decreases during the decay phase \citep{Campana2014}, the bolometric correction factor also increases. For example, the gravitational redshift-corrected effective temperature at the onset of the sharp decay ($1\times10^{37}~$\ergpers) is approximately $300~$eV \citep{Gungor2014}, whereas during quiescence ($5\times{10}^{33}~$\ergpers), it drops to $120~$eV. We have estimated that the fraction of the total blackbody flux emitted in the observational band decreases from about $88\%$ to $38\%$. This change does not significantly change the morphology of the light curve.

The sharp decay in $\Lx$ starts when $\rh$ decreases to about $10^{10}~\mathrm{cm}$. After the knee, the model cannot account for the functional form of the sharp decay curve without including the local $\Mdotin$ dependence of the $h/r$ ratio and its effect on $\Firr$. 
We smooth out the realistic opacities to prevent numerical fluctuations. The $\Lx$ fluctuations during the quiescent state are due to continuous mass flow toward the inner region of the cold outer disk and the resultant local instabilities continually taking part of this region into the hot state. We calculate the entire $\Lx$ curve with $\Mdotstar = \Mdotin$. That is, $\Lx$ is produced by accretion onto the star without invoking any partial accretion or strong propeller. In this case, we can estimate an upper limit to the dipole field strength $B$ on the surface of the star. 
Using equation~(\ref{eqn:Rin}) and the strong propeller condition $\reta > 1.26 \rco$ (see \citealt{Ertan2021} for details), the $\Mdotin$ level in the quiescent state ($\sim 1.5 \times 10^{14}~$\gpers) gives $B < 4 \times 10^8$ G.

A model including the $\Mdotin$ dependence of the disk aspect ratio in the $\Firr$ calculation produces a much sharper decay curve after the knee in comparison with a model neglecting this dependence. Figure~\ref{fig:1} shows the difference in the model curves obtained for these two cases.

How does a very weak $\Mdotin$ and $r$ dependence of $h/r$ yield the difference in $\Lx$ curves seen in Figure~\ref{fig:1}? This can be understood by comparing hot disk radii, $\rh'$ and $\rh$, obtained with constant and varying $h/r$, respectively. From equation~\ref{Eqn:Firr}, it can be shown that $\rh' \propto \Mdotin^{1/2}$. Using C given in equation~\ref{C}, we obtain $\rh \propto \Mdotin^{46/75}$. During the decay phase, both $\rh$ and $\rh'$ decrease, while the $\rh'/\rh$ ratio increases. Comparing the model curves in the bottom and upper panels of Figure~\ref{fig:1}, it is seen that $\rh \simeq \rh'$ when $\Lx \simeq 10^{37}~$\ergpers. With two orders decrease in $\Lx$ to $\sim 10^{35}~$\ergpers, $\rh'/\rh$ increases to $1.7$. 

It is the dynamics of the hot inner disk that determines the X-ray outburst light curve morphology. The matter at the outermost regions of the hot disk contributes to accretion onto the star after one viscous time across the disk. Relatively fast inward propagation of $\rh$ also reduces the amount of hot matter more rapidly, which produces a cumulative effect leading to a sharper decline in $\Lx$, and thus to an even faster shrinking size of the hot disk. For the case with constant $C$, the disk evolves with a larger hot disk size, that is, more matter flows to the star during the outburst state. This produces an outburst with relatively high $\Lx$ and long duration. 

The model agrees with the observed $\Lx$ level in quiescence as well. The observed $\Lx$ is not smooth in quiescence. There are flares with ten-fold variations \citep{CotiZelati2014,Bernardini2013}.
These $\Lx$ fluctuations are also seen in our model curve. Nevertheless, correct modeling of this behavior is not easy due to the resolution problem for the innermost disk for a disk size greater than $10^{11}$ cm, which is beyond the scope of this work.   
We should also note that there are differences in the observed $\Lx$ levels after different outbursts \citep{Campana2014}. 

\subsection{The Quiescent State and Recurrent Outbursts}

\begin{figure}[ht!]
\centering
\includegraphics[width=\columnwidth]{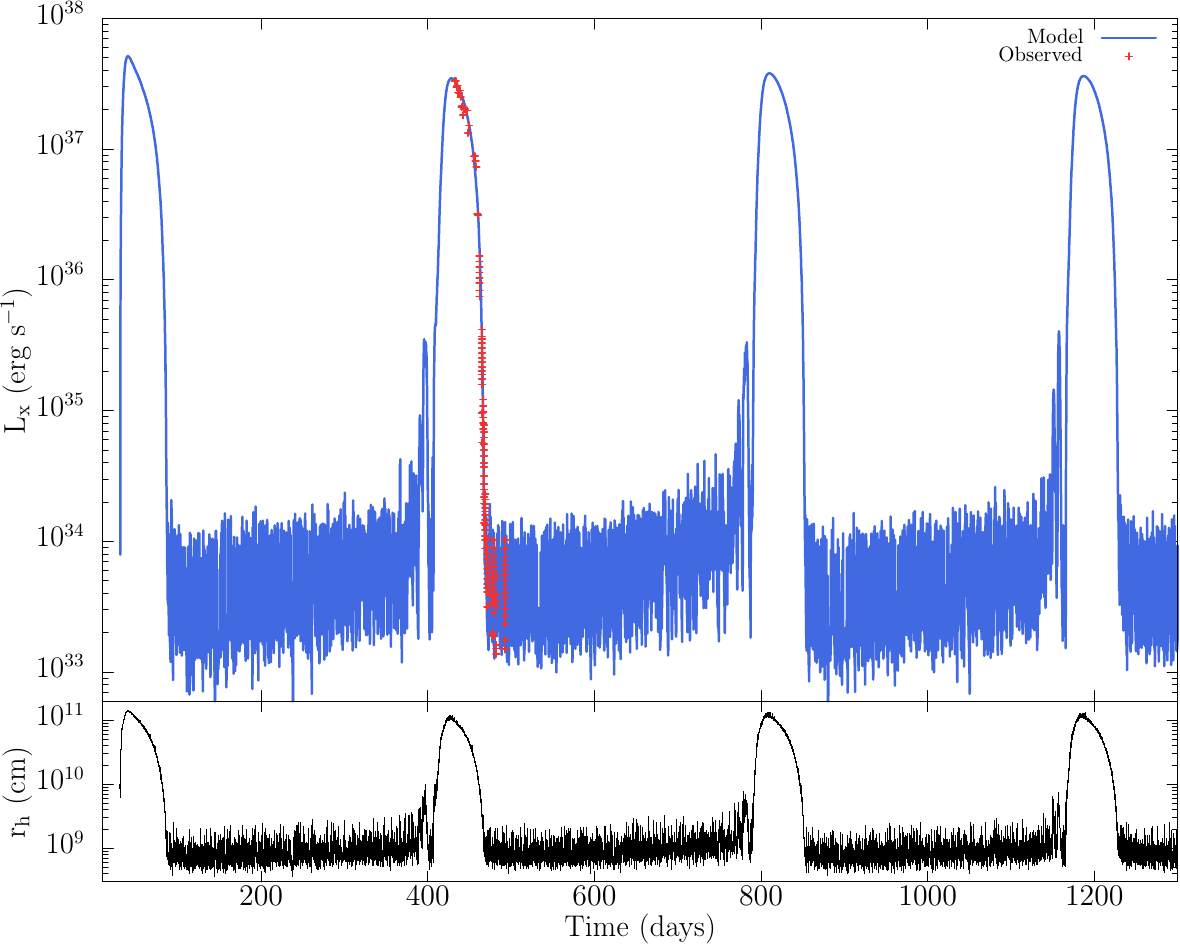}
\caption{The same model parameters are used for the solid curve in Figure~\ref{fig:1}. This model curve is obtained by injecting mass into the radial grid at $r= \rcirc$ with a rate of $4 \times 10^{16}~$\gpers. The recurrence time of the outbursts is approximately one year. The second outburst seen in this figure is enlarged in Figure~\ref{fig:3}. The bottom panel shows the evolution of the hot disk radius $\rh$.}
\label{fig:2}
\end{figure}

\begin{figure}[ht!]
\centering
\includegraphics[width=\columnwidth]{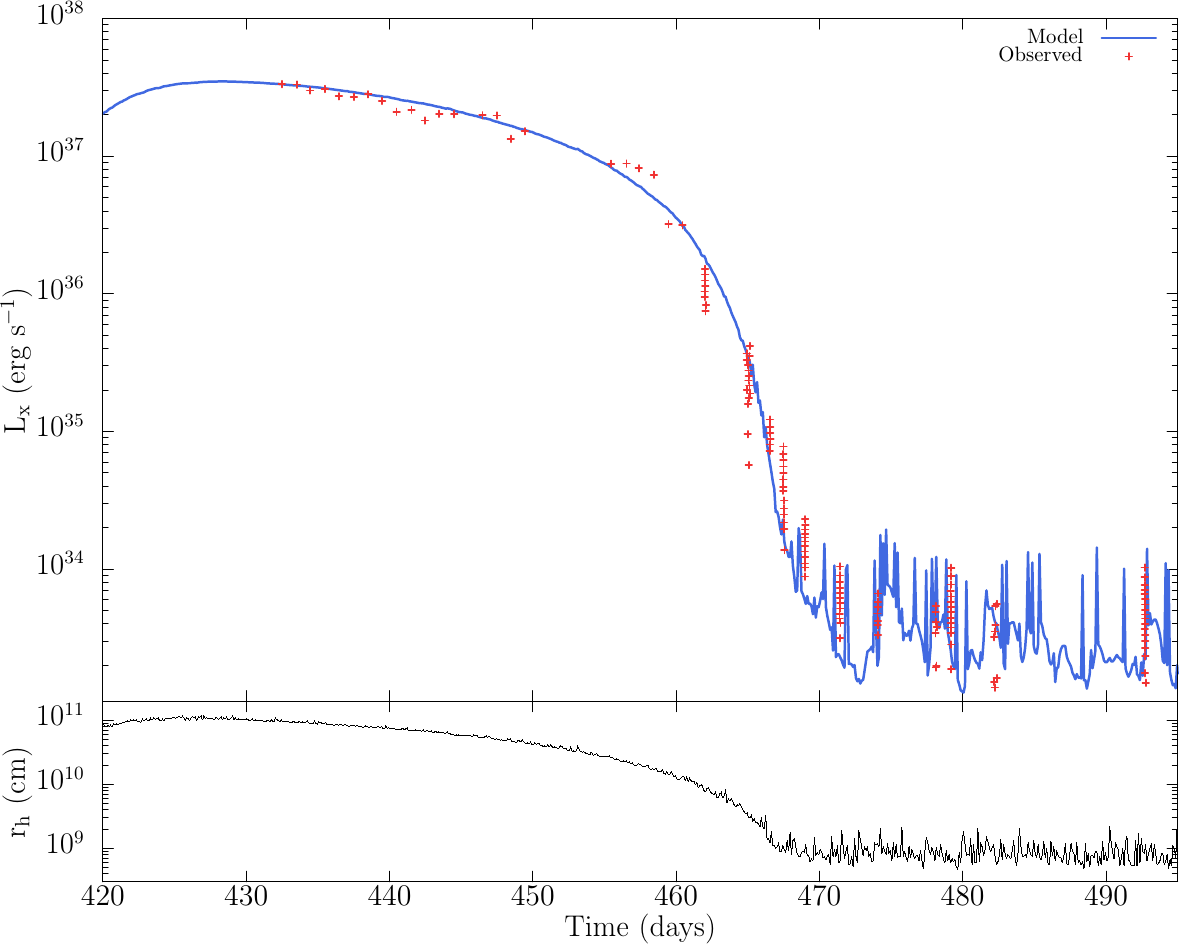}
\caption{The second outburst seen in Figure~\ref{fig:2}. The bottom panel shows the evolution of the hot disk radius $\rh$.}
\label{fig:3}
\end{figure}

Using the same model parameters producing the model curve in Figure~\ref{fig:1}, we have continuously added mass into the radial grid at $\rcirc = 5 \times 10^{10}$ cm to represent the mass flow into the outer disk from the companion. The outburst light curves seen in Figure~\ref{fig:2} are obtained with $\Mdotc = 4 \times 10^{16}~$\gpers~which produces outbursts with $\Trec \sim 1$ year. It is interesting that the model light curves of the second and fourth outbursts also seemingly fit well to the X-ray outburst data of Aql X-1. For instance, Figure~\ref{fig:3} zooms in on the second outburst seen in Figure~\ref{fig:2}. 

\begin{figure}[ht!]
\centering
\includegraphics[width=\columnwidth]{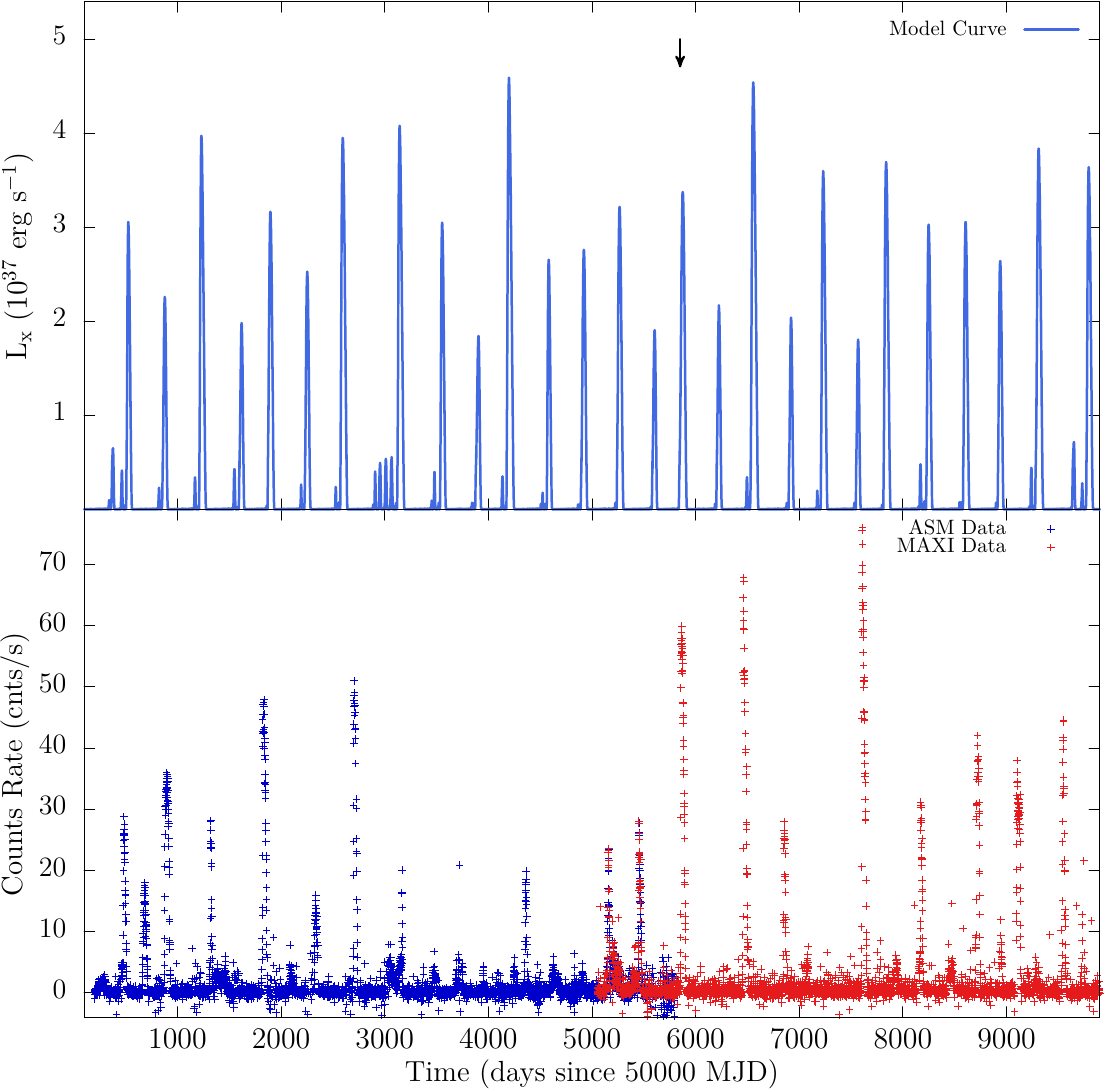}
\caption{Top Panel: Long-term model light curve produced with $\alphaC = 7.5\times10^{-3}$ and $\Mdotc=4 \times 10^{16}~$\gpers. The other parameters are the same as those given in Figure~\ref{fig:1}. Bottom Panel: The Long-term light curve of Aql X-1 from May 1996 to January 2023 observed by ASM(Blue) and MAXI(Red).}
\label{fig:4}
\end{figure}

\begin{figure}[ht!]
\centering
\includegraphics[width=\columnwidth]{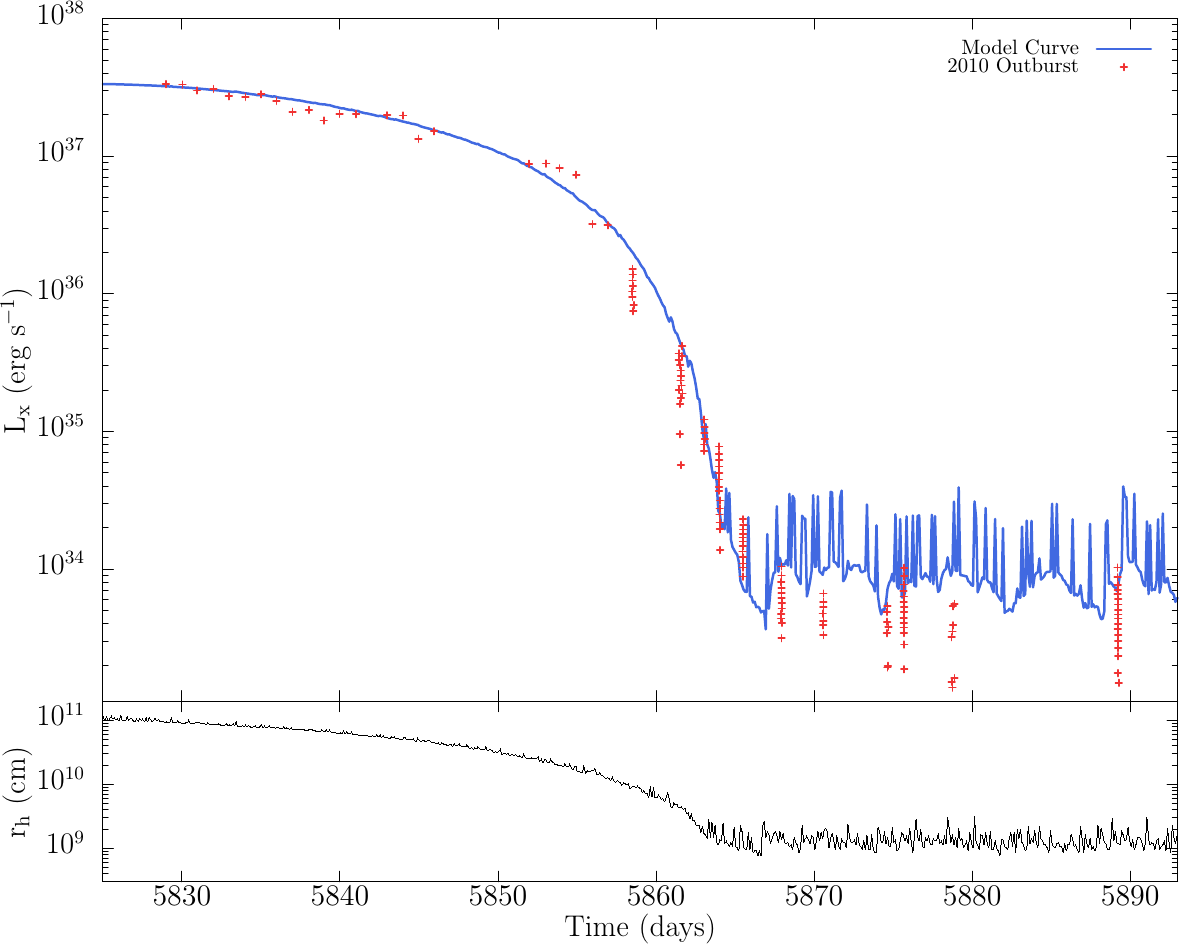}
\caption{Enlarged model outburst curve shown by the arrow in Figure~\ref{fig:4} (bolometric $\Lx$). Red data points show the $\Lx$ curve estimated from the observed $0.5 - 10$ keV flux data during the 2010 outburst of Aql X-1 for $d = 4.5$ kpc \citep{Campana2014}. The bottom panel shows the evolution of the hot disk radius $\rh$.}
\label{fig:5}
\end{figure}

For a comparison of our long-term ($\sim 10^{4}$ days) model light curve with the observations, we use the $2-10$ keV data from the \textit{All-Sky Monitor} (ASM) on the \textit{Rossi X-ray Timing Explorer} and the \textit{Monitor of All-sky X-ray Image} (MAXI) instrument mounted on the International Space Station (ISS). Count rates from these detectors are calibrated using the 2009 and 2010 outbursts which were observed by both instruments. In Figure~\ref{fig:4}, it is seen that the relative amplitudes and the recurrence time-scale of the outbursts produced in the model are very similar to the observed data. In the model, the recurrence time, relative amplitudes, and morphologies of the outbursts as well as the quiescent $\Lx$ level are sensitively depend on the $\alphaC$ and $\Mdotc$ parameters. The model curve in Figure~\ref{fig:4} is obtained with $\alphaC = 7.5\times10^{-3}$ and $\Mdotc = 4 \times 10^{16}~$\gpers. Relatively high $\alphaC$ produces outburst curves with higher quiescent $\Lx$ levels and relative amplitudes. For a given $\alphaC$, higher values of $\Mdotc$ yield outbursts with smaller $\Trec$. For instance, the model curves seen in Figures~\ref{fig:3} and \ref{fig:5} are obtained with $\alphaC$ values of $5\times10^{-3}$ and $7.5\times10^{-3}$ respectively. It is seen that the $\Lx$ level in the quiescent phase is a few times higher for the model with greater $\alphaC$. The data from the 2010 outburst were obtained in the energy range of $0.5-10$ keV without bolometric correction. The blackbody temperature estimated from the spectral fits is about $120$ eV in the quiescent state \citep{Campana2014}. This indicates a bolometric correction factor of $\sim 2.7$ in the quiescence, which means that the model with $\alphaC = 7.5\times10^{-3}$ (Figure~\ref{fig:5}) is in better agreement with the corrected $\Lx$ level in the quiescent state.

The light curve morphologies of the recurrent outbursts seen in Figures~\ref{fig:2} and \ref{fig:4} are not sensitive to the initial mass distribution in the model. Like the model with $\alphaC = 5\times10^{-3}$, a fraction of the outburst model curves with $\alphaC = 7.5\times10^{-3}$ naturally yields the morphology of the observed 2010 outburst light curve of Aql X-1. For instance, the model outburst given in Figure~\ref{fig:5} is the same outburst indicated by an arrow in Figure~\ref{fig:4}. Indeed, there are observed outburst light curves of the source very similar to that of the 2010 outburst.
    
In the quiescent state of Aql X-1, $\Lx \gtrsim 5 \times 10^{33}~$\ergpers. For comparison, tMSPs are estimated to accrete matter from the disk down to $\Lx$ levels of $\sim 10^{33}~$\ergpers~\citep{archibald2009,Papitto2022,bassa2014state}. Below this level, X-ray pulses are switched off, and they show RMSP behavior. Dipole field strengths of tMSPs seem to be similar to that indicated by the upper limit $B < 4 \times 10^8$ G  estimated for Aql X-1 in this work. Considering also the similarities in the rotational properties, Aql X-1 could have a critical $\Mdotin$ rate for termination of accretion and transition to the propeller phase that is similar to that of tMSPs. This might be the reason for the lack of radio pulsations from Aql X-1 during its quiescent state. It is important to note here that the minimum 
pulsed $\Lx$ level of tMSPs is about two orders of magnitude smaller than the rate corresponding to $\rxi = \rco$.

\section{Conclusions}
\label{sec:Conclusions}
We have shown that the characteristic X-ray outburst light curve features of Aql X-1 can be reproduced by the mass accretion onto the neutron star along the entire X-ray outburst phase without requiring a transition into the propeller phase. The knee at the end of the initial slow exponential decay comes about as the cold/hot border of the disk approaches the inner disk, which is in line with the results obtained by \citet{Lipunova2022}. 
The sharp decay phase after the knee can be accounted for by taking the weak $\Mdotin$ and $r$ dependence of the $h/r$ ratio into account in the $\Firr$ calculation. 
Throughout the outburst phase, we take $\Mdotin = \Mdotacc$ without including any partial accretion regime or propeller effect in the $\Lx$ calculations. Injecting mass into the grid at $\rcirc$ with a rate of $\sim 4 \times 10^{16}~$\gpers, the model can also generate recurrent X-ray outbursts that mimic typical outburst light curves of Aql X-1 consistently with the observed $\Trec \sim 1$ yr. We have also estimated that $B < 4 \times 10^{8}$ G corresponding to $\rin \leq \rco$ for the lowest $\Lx$ level of the source.

\begin{acknowledgements}
We acknowledge research support from T\"{U}B{\.I}TAK (The Scientific and Technological Research Council of Turkey) through grant 120F329 and from Sabanc\i\ University. We thank Ali Alpar for the useful comments that improved our manuscript. This research has made use of MAXI data provided by RIKEN, JAXA, and the MAXI team \citep{MAXIPaper} as well as results provided by the ASM/RXTE team.
\end{acknowledgements}

\bibliography{AqlX-1}{}
\bibliographystyle{aasjournal}



\end{document}